# Very Large Tunneling Magnetoresistance in Layered Magnetic Semiconductor $CrI_3$

Zhe Wang[1,2*], Ignacio Gutiérrez-Lezama[1,2], Nicolas Ubrig[1,2], Martin Kroner[3], Marco Gibertini[1,2], Takashi Taniguchi[4], Kenji Watanabe[4], Ataç Imamoğlu[3], Enrico Giannini[1] and Alberto F. Morpurgo[1,2*]

[1]*Department of Quantum Matter Physics, University of Geneva, 24 Quai Ernest Ansermet, CH-1211 Geneva, Switzerland*

[2]*Group of Applied Physics, University of Geneva, 24 Quai Ernest Ansermet, CH-1211 Geneva, Switzerland*

[3]*Institute of Quantum Electronics, ETH Zürich, CH-8093 Zürich, Switzerland*

[4]*National Institute for Materials Science, 1-1 Namiki, Tsukuba 305-0044, Japan*

[*]email: zhe.wang@unige.ch   Alberto.Morpurgo@unige.ch

## Abstract

**Magnetic layered van der Waals crystals are an emerging class of materials giving access to new physical phenomena, as illustrated by the recent observation of 2D ferromagnetism in $Cr_2Ge_2Te_6$ and $CrI_3$. Of particular interest in semiconductors is the interplay between magnetism and transport, which has remained unexplored. Here we report first magneto-transport measurements on exfoliated $CrI_3$ crystals. We find that tunneling conduction in the direction perpendicular to the crystalline planes exhibits a magnetoresistance as large as 10'000 %. The evolution of the magnetoresistance with magnetic field and temperature reveals that the phenomenon originates from multiple transitions to different magnetic states, whose possible microscopic nature is discussed on the basis of all existing experimental observations. This observed dependence of the conductance of a tunnel barrier on its magnetic state is a new phenomenon that demonstrates the presence of a strong coupling between transport and magnetism in magnetic van der Waals semiconductors.**

## Introduction

Investigations of layered van der Waals compounds are revealing a wealth of electronic phenomena, which can be controlled by varying the material thickness at the atomic scale[1-4]. Among these compounds, magnetic van der Waals semiconductors[5-19] have remained virtually unexplored. These materials possess a unique potential for new physical phenomena, because magnetism occurs spontaneously without the need to introduce magnetic dopants as done in conventional magnetic semiconductors[20-22], allowing –at least in principle– perfect crystalline order to be preserved. Indeed, the potential of magnetic van der Waals semiconductors has been made apparent by very recent experiments showing the occurrence of 2D ferromagnetism in atomically thin layers of $Cr_2Ge_2Te_6$ [16] and $CrI_3$ [17]. So far however, essentially no experiment has been done to probe the transport and opto-electronic properties of these materials, and it remains to be determined whether their behavior deviates from that of conventional semiconductors, i.e., whether magnetism causes new interesting physical phenomena to appear. This can be expected because *ab-initio* calculations predict the valence and conduction band of several ferromagnetic van der Waals semiconductors to be fully spin polarized[15, 23-26], implying a very strong coupling between the magnetic state and other electronic properties. Here we investigate experimentally these issues by performing transport and optical measurements on nano-fabricated devices based on exfoliated $CrI_3$ crystals. In all devices investigated we find a very large tunneling magnetoresistance –as large as 10'000% –originating from abrupt transitions between different magnetic states of $CrI_3$ that shows directly how the transport and magnetic properties are strongly coupled.

## Results

**Semiconducting characteristics of $CrI_3$ devices**. Past studies[5, 6, 10] have shown that $CrI_3$ exhibits a transition to an anisotropic ferromagnetic state with easy axis perpendicular to the layers (Curie temperature $T_c$ =61 K), accompanied by a singular behavior of the magnetic susceptibility below $T \sim 50$ K suggestive of a transition to a more complex magnetic stat that remains to be understood (see Supplementary Note 2). In contrast to the magnetic properties, virtually nothing is known about the opto-electronic response of this material and –to start exploring it– we have fabricated and investigated different types of devices (Fig. 1a-d, see Methods section and Supplementary Note 3 for details of the fabrication process). Fig. 1 a and b show the schematics and an optical microscope image of a structure with graphene contacts attached to the bottom of an exfoliated $CrI_3$ crystal, that we realized to implement a field-effect transistor (the doped Si substrate acts as gate). The observed gate and bias dependence of the current are shown in Fig. 1e and its inset: they conform to the expected transistor behavior and indicate that transport in $CrI_3$ is mediated by electrons in the conduction band (since the transistor turns on at positive gate voltages). Fig. 1c-d show a second type of devices with contacts connected on opposite sides of an exfoliated thin $CrI_3$ crystal, enabling photocurrent measurements. The photocurrent sets in sharply when the photon energy exceeds 1.2 eV (Fig. 1f), corresponding to the $CrI_3$ band gap[5, 6]. This value is consistent with that inferred from the $CrI_3$ photoluminescence spectrum that peaks at the same energy (see red line in Fig. 1f). Notably, the magnitude of the photocurrent is comparable to that measured on analogous devices based on crystals of "more established" van der Waals semiconductors such as WS$_2$ [27] or WSe$_2$ [28]. In contrast to the field effect transistors, whose resistance was found in all cases to become unmeasurably high below 100 K, the photocurrent in vertical junctions persists down to low temperature. This suggests that the measurement of "vertical transport" in the direction perpendicular to the layers is possible at low temperature, and may be used to probe phenomena of magnetic origin.

**Tunneling transport in vertical junctions**. We investigated "vertical transport" from room temperature down to $T = 0.25$ K, by measuring the *I-V* curves of devices such as the one in Fig. 1d. Representative data from one of these devices (Fig. 2a, thickness of CrI$_3$ is ~ 7 nm, corresponding to approximately 10 monolayers) show strongly non-linear *I-V* curves that are temperature independent for $T < 20$ K, whereas for larger *T* the current *I* at any given bias *V* increases with increasing temperature. The temperature evolution of the resistance *R* (=*V*/*I*) measured at three different biases (*V*= 0.35 V, 0.5 V, and 0.7 V) is summarized in Fig. 2b: starting from room temperature, the resistance first increases in a thermally activated way down to $T \sim 70$ K (the typical value of activation energy found in different devices is $E_a$ ~0.15 eV), where it starts to level off, and eventually saturates becoming temperature independent for $T < 20$ K.

The observed temperature independence indicates that for $T < 20$ K vertical transport is due to tunneling. Indeed, Fig. 2c shows that, for $T < 20$ K, $ln(I/V^2)$ scales proportionally to $1/V$, the trend expected in the Fowler-Nordheim (FN) tunneling regime[29, 30]. This regime occurs when the electric field generated by the applied voltage tilts the bands in the semiconductor, allowing carriers to tunnel from the electrode into the material[31] (see the inset of Fig. 2c). Increasing the electric field effectively decreases the barrier thickness and causes an exponential increase of current. Theory predicts:

$$ln\frac{I}{V^2} \sim -\frac{8\pi\sqrt{2m^*}\phi_B^{3/2}d}{3hqV}, \qquad (1)$$

where *h* is Planck's constant, *q* the electron charge, *d* the barrier thickness, *m*\* the effective mass and $\phi_B$ the barrier height determined by the distance between the Fermi energy in the contact and the edge of the conduction band in CrI$_3$ (the transistor measurements in Fig. 1e imply that electrons –and not holes– are responsible for the "vertical" tunneling current). If the effective mass is taken to be equal to the free electron mass –a plausible assumption in view of the rather narrow bands of CrI$_3$[10, 23, 24]– we find that the barrier height is 0.25 eV, roughly comparable to the activation energy extracted from the measured temperature dependence of the resistance.

**Large tunneling magnetoresistance in vertical junctions**. Having established the mechanism of vertical transport and seen that measurements can be done well below the Curie temperature, we look at the effect of an applied magnetic field. The magnetoresistance measured at different temperatures between 10 K and 65 K with the magnetic field applied perpendicular to the plane of CrI$_3$ is shown in Fig. 3a-f (extra data are discussed in Supplementary Note 4 and 5). Extremely large "jumps" are observed at low temperature, resulting in a total magnitude change up to 10'000% as **B** is increased from 0 to just above 2 T (Fig. 2a). The jumps pointed by the vertical arrows (J1, J2, and J3) are seen in all four measured devices. "Jumps" J2 and J3 occur at the same values of **B** irrespective of the CrI$_3$ thickness, which in our experiments ranged from 5.5 nm to 14 nm (additional fine structure in the data depends on the specific device measured). Such a large magnetoresistance is striking as it is not commonly observed for electrons tunneling through non-magnetic materials. The sharp and well-defined values of applied magnetic field at which the resistance jumps are seen strongly suggest that the phenomenon originates from changes in the magnetic state of the CrI$_3$ layers.

**Magnetic states of CrI$_3$ and tunneling magnetoresistance**. Identifying the nature of the magnetic states responsible for the tunneling magnetoresistance, and checking consistency with the known magnetic properties of bulk CrI$_3$ is subtle. We start addressing these issues by investigating how the magnetoresistance depends on temperature. Upon increasing $T$, the resistance jumps shift position (compare Fig. 3a and 3c), with J2 and J3 becoming less sharp (Fig. 3d), and all features eventually disappear around 50 K, well below the Curie temperature of CrI$_3$ ($T_c$ = 61 K) at which the magnetization of the material appears. More detailed information is obtained by looking at the dependence of the magnetoresistance on both temperature and magnetic field. The color plot in Fig. 4a clearly shows that the resistance "jumps" define three states (that we label as I, II, and III). For $T$ lower than approximately 40 K the states are separated by clear boundaries in the **B**-$T$ plane, and well-defined transitions are seen irrespective of whether the boundary is crossed by varying $B$ at fixed $T$ or by changing $T$ at fixed **B** (as shown in Fig. 4b), as expected for veritable phase transitions. This confirms that different magnetic states in CrI$_3$ are responsible for the observed magnetoresistance.

Determining the precise onset temperature for the occurrence of tunneling magnetoresistance is also instructive. For $T > 40$ K the jumps are rounded into "kinks" whose position can be determined as shown in Fig. 4c. By following their evolution in the $B$- and $T$-plane we find that jump J3 (see the red circles in Fig. 4a) and J1 (see Fig. 4c, the feature associated to the small field hysteretic behavior visible in Fig. 3) start at the same temperature, $T \cong 51$ K (rounding prevents the precise evolution of jump J2 to be followed above 40 K). Notably, 51 K corresponds exactly to the temperature of the singular behavior observed in the magnetic susceptibility of bulk crystals (see Supplementary Fig. 1), indicative of a transition to a complex magnetic state different from a simple ferromagnet. This quantitative agreement therefore suggests that one of the states responsible for the large tunneling magnetoresistance observed in the experiments is the same state that manifests itself in the magnetic properties of bulk CrI$_3$. The relation between the properties of bulk crystals and magnetoresistance is however more complex, as shown by magneto-optical Kerr effect (MOKE) measurements.

MOKE measurements exhibit a behavior analogous to that observed in transport, with sharp jumps in Kerr angle that are seen upon the application of a magnetic field perpendicular to the plane of $CrI_3$ layers (Faraday geometry[32]; see Fig. 5a). The jumps occur precisely at the same **B**-values at which the J2 and J3 magnetoresistance jumps are found. At $T = 5$ K a well-developed hysteresis in the magnetoresistance measurements is observed upon sweeping **B** up and down, and the same hysteresis is seen in the measurements of Kerr angle. The evolution of the $B$-dependence of the Kerr angle upon increasing $T$ (see Fig. 5b-c) also resembles what is observed in the magnetoresistance, with the jumps in the two quantities shifting and smearing in a very similar way. All the trends that we observed are qualitatively identical to the one reported earlier in atomically thin layers[17], which –after submission of this manuscript– have also been reported to exhibit a tunneling magnetoresistance virtually identical to the one reported here[33, 34].

In thin atomic layers, MOKE measurements have been interpreted in terms of individual crystalline planes of $CrI_3$, antiferromagnetically coupled at **B** = 0, switching to a ferromagnetic ordering upon application of an external magnetic field. Switching from antiferro- to ferro-magnetic coupling may account for the occurrence of sharp jumps in the tunneling magnetoresistance, but is at odds with bulk magnetic properties. Indeed, if transitions from antiferromagnetic to ferromagnetic ordering of $CrI_3$ layers would occur in crystals of all thicknesses, very large jumps should be observed in the bulk magnetization. Direct measurements (Supplementary Fig. 1) however show a near complete saturation of bulk magnetization occurring already at **B** ~ 0.3 T, and no change at **B** ~ 0.9 T and ~ 1.8 T in correspondence of the magnetoresistance jumps.

We conclude that the comparison of different experiments ($T$- and **B**-dependence of the resistance, magnetic susceptibility, MOKE) indicate that the tunneling magnetoresistance originates from transitions in the magnetic state of $CrI_3$. As for the details of the magnetic states involved, however, no simple interpretation straightforwardly reconciles all observations made on both bulk crystals and thin exfoliated layers. More work is needed to clarify this issue and two potentially important points are worth mentioning. One is that the evidence for the possible antiferromagnetic coupling proposed for exfoliated layers is inferred from the MOKE measurements, under the assumption of a direct relation between Kerr angle and total magnetization. However, it is known that in semiconducting systems a finite Kerr effect can be observed in the absence of a net magnetization as long as time reversal and inversion symmetry are broken[35, 36]. Hence, transitions in the magnetic state without any change in magnetization (see illustrative examples in Supplementary Fig. 8 and discussion in Supplementary Note 6) could –at least in principle–cause jumps in Kerr rotation. This implies that to establish conclusively that an antiferromagnetic coupling is present it is very important to measure the magnetization of atomically thin $CrI_3$ layers directly without simply relying on MOKE.

A second point to be made is that the jumps observed in MOKE and magnetoresistance could indeed originate from an antiferromagnetic coupling of adjacent $CrI_3$ layers, but that inter-layer antiferromagnetism only occurs in an interfacial region close to the crystal surface. If sufficiently thin, such an interfacial region would not be detected in bulk magnetization measurements, explaining why no magnetization jump is observed at 0.9 T and 1.8 T. Our observations that magnetoresistance occurs always at the same magnetic field values irrespective of applied bias, polarity, and thickness of the $CrI_3$ flake up to 20 monolayers indicate that if an interfacial region is invoked, this region is rather thick. We estimate that crystals between 10 and 20 monolayers thick are still fully antiferromagnetically coupled, and understanding why such thick crystals exhibit a behavior that very different from the one observed in the bulk is not obvious. One possibility is that the crystal structure of exfoliated layers in the surface region is different from that of bulk crystals. Interestingly, *ab-initio* calculations show that if individual layers are stacked according to the high-temperature crystalline phase of $CrI_3$, an antiferromagnetic interlayer coupling is energetically favorable as compared to the ferromagnetic coupling found in the low-temperature crystalline structure (see Supplementary Fig. 9 and discussion in Supplementary Note 7).

**Coupling between magnetic state and tunneling resistance**. Irrespective of the microscopic details of the underlying magnetic structure, finding that a step-like large modulation in the tunneling resistance of a magnetic insulator can be induced by a change in its magnetic state is a physical phenomenon that has not been reported earlier. It is therefore useful to look in detail at the *I-V* curves of our devices to see whether the experiments provide any indication as to the microscopic mechanism responsible for the change in tunneling resistance in the different magnetic states. This is done in Fig. 6a where $ln(I/V^2)$ is plotted versus $1/V$ for many different values of magnetic field. It is apparent that for different magnetic field intervals (I: 0-0.9 T; II: 1-1.8 T; III: 1.9-3 T) the I-V curves collapse on top of each other. In all cases the overall behavior is consistent with that expected from FN tunneling, but with a constant of proportionality between $ln(I/V^2)$ and $1/V$ (i.e. the slope of the three dashed lines in Fig. 6a) that is different in the three cases.

Microscopically, the constant of proportionality between $ln(I/V^2)$ and $1/V$ is determined by the transmission probability, i.e. to the extinction of the electron wavefunction tunneling through the $CrI_3$ barrier. Within the simplest model of a uniform (i.e. layer independent) gap of $CrI_3$, this quantity is determined by the electron effective mass and by the height of the tunnel barrier (see Eq. (1)). Using Eq. (1) under the assumption that the effective mass remains unchanged, we can extract the height of the tunnel barrier from the slope of the $ln(I/V^2)$ versus $1/V$ relation. We find that –as shown in Fig. 6b – the barrier height is different in different magnetic states. The effect can be due to either a change in the band-gap or in the work-function of $CrI_3$, consistently with *ab-initio* calculations which shows that the occurrence of magnetism is accompanied by a modification in the material band structure[10, 23, 24].

A similar conclusion –namely that different magnetic states have different transmission probability because of a difference in tunnel barrier height– holds true even if the change in tunneling magnetoresistance is due to a switch from an antiferromagnetic to a ferromagnetic interlayer ordering of the magnetization. The simplest model to describe this scenario consists in assuming that each layer has different barrier height for spin up and down[37, 38] (i.e., the height of the tunnel barrier is not spatially uniform). Calculations based on this model lead to *I-V* curves that also approximately satisfy Fowler-Nordheim behavior, with proportionality constant between $ln(I/V^2)$ versus $1/V$ that is different for a ferromagnetic or antiferromagnetic alignment of the magnetization in the individual $CrI_3$ layers (See Supplementary Fig. 10). This behavior is easy to understand qualitatively, because –despite not being spatially uniform– the average height of the tunnel barrier for the tunneling process that gives the dominant contribution to the current also depends on the magnetic state. Specifically, for ferromagnetic alignment, tunneling is dominated by the majority spin and the barrier height is the same –the smallest possible– in all layers. For antiferromagnetic coupling, electrons experience a barrier height that is alternating (depending on the layer) between the value expected for majority and minority spins, larger on average that the height experienced by the majority spins in the case of ferromagnetic alignment. These considerations imply that –at least at the simplest level– the analysis of the FN tunneling regime in the measured *I-V* curves cannot discriminate between different magnetic states. Nevertheless, they also do indicate that the magnetic state is coupled to the band structure, which is why the height of the tunnel barrier depends on the specific magnetic state.

## Discussion

The observation that the tunneling conductance through a $CrI_3$ barrier depends strongly on the magnetic state of the material –a phenomenon that had not be observed previously in other systems– showcases the richness of physical phenomena hosted by van der Waals semiconductors. So far, these materials have attracted attention mainly for their opto-electronic and transport properties, but it is becoming apparent that their magneto-electronic response also exhibits fascinating and possibly unique properties. Inasmuch $CrI_3$ is concerned, future experiments should identify which other electronic phenomena, besides the tunneling conductance, are strongly affected by the magnetic state of the material. They should also aim at improving the quality of field-effect transistors (whose operation for $CrI_3$ has been demonstrated here) to enable the investigation of gate-controlled transport at low temperature, in the magnetic state of the material. Obviously, however, experiments should be performed on a broader class of van der Waals magnetic systems, starting with those for which past measurements of the bulk magnetic response indicate the occurrence of transitions between states that can be controlled by the application of an experimentally reachable magnetic field.

After the submission of our manuscript different preprints have appeared on the cond-mat archive reporting observations closely related to the ones discussed here (Refs. 33,34,39).

## Methods:

**Crystal growth**: High quality crystals of $CrI_3$ have been grown by the chemical vapor transport method in a horizontal gradient tubular furnace. To avoid degradation of the precursors and synthesized crystals the 1:3 mixture of Cr and I was sealed in a quartz tube (later placed in the furnace) under inert conditions (see Supplementary Note 1).

**Sample fabrication**: Multilayer graphene, h-BN (10 – 30 nm) and few-layer $CrI_3$ flakes were exfoliated in a nitrogen gas filled glove box with a < 0.5 ppm concentration of oxygen and water to avoid degradation of the few-layer $CrI_3$ crystals, which are very sensitive to ambient conditions (see Supplementary Note 3). The heterostructures were then assembled in the same glove box with a conventional "pick-up and release" technique based on either PPC/PDMS or PC/PDMS polymer stacks placed on glass slides. Once encapsulated, the multilayer graphene electrodes were contacted electrically by etching the heterostructures by means of reactive ion etching (in a plasma of a $CF_4/O_2$ mixture) followed by evaporation of a 10nm/50nm Cr/Au thin film.

**Transport measurements**: Transport measurements were performed either in a Heliox $^3$He insert system (Oxford Instruments, base temperature of 0.25 K) equipped with a 14 T superconducting magnet, or in the variable temperature insert of a cryofree Teslatron cryostat (Oxford Instruments, base temperature of 1.5 K) equipped with a 12 T superconducting magnet. The latter system is also equipped with a sample rotator making it possible to align the sample so that the magnetic field is either parallel or perpendicular to the $CrI_3$ layers. The *I-V* curves and magneto-resistance were measured with a Keithley 2400 source/measure unit and/or home-made low-noise voltage sources and current amplifiers.

**Optical measurements & MOKE**: Photoluminescence measurements were performed in a home tailored confocal micro-photoluminescence setup in back-scattering geometry (i.e. collecting the emitted light with the same microscope used to couple the laser beam onto the device). The light collected from the sample was sent to a Czerny-Turner monochromator and detected with a liquid nitrogen cooled Si CCD-array (Andor emCCD). The sample was illuminated with the 647.1 nm laser line of an Ar-Kr laser at a power of 30 μW. The data were corrected to account for the non-linear CCD response in this spectral region. The same setup was used for the photocurrent measurement but in this case the devices were illuminated using a Fianium supercontinuum laser coupled to a monochromator, providing a beam of tunable wavelength with spectral width of 2 nm and stabilized power.

The magneto-optical Kerr effect (MOKE) measurements were performed in a cryostat with a 12 T superconducting split-coil magnet. The sample was illuminated with linear polarized light at 632.8 nm and 50 μW provided by a power stabilized HeNe laser. The reflected beam was split using a polarizing beam splitter cube and the s- and p-components measured simultaneously with Si-photodiodes and lock-in detection. A linear contribution to the measured Kerr angle, which stems from Faraday rotation in the cold objective lens, has been measured independently and was subtracted from the data in order to obtain the traces shown in Fig. 4.

**Data availability**: All relevant data are available from the corresponding authors on request.

## References:

1. Castro Neto A. H., Guinea F., Peres N. M. R., Novoselov K. S. & Geim A. K. The electronic properties of graphene. *Rev. Mod. Phys.* **81,** 109-162 (2009).

2. Wang Q. H., Kalantar-Zadeh K., Kis A., Coleman J. N. & Strano M. S. Electronics and optoelectronics of two-dimensional transition metal dichalcogenides. *Nat. Nanotech.* **7,** 699-712 (2012).

3. Xu X., Yao W., Xiao D. & Heinz T. F. Spin and pseudospins in layered transition metal dichalcogenides. *Nat. Phys.* **10,** 343-350 (2014).

4. Novoselov K. S., Mishchenko A., Carvalho A. & Castro Neto A. H. 2D materials and van der Waals heterostructures. *Science* **353,** (2016).

5. Dillon J. F. & Olson C. E. Magnetization Resonance and Optical Properties of Ferromagnet $CrI_3$. *J. Appl. Phys.* **36,** 1259-1260 (1965).

6. Dillon J. F., Kamimura H. & Remeika J. P. Magneto-optical properties of ferromagnetic chromium trihalides. *J. Phys. Chem. Solids* **27,** 1531-1549 (1966).

7.  Carteaux V., Moussa F. & Spiesser M. 2D Ising-Like Ferromagnetic Behavior for the Lamellar $Cr_2Si_2Te_6$ Compound: A Neutron-Scattering Investigation. *Europhys. Lett.* **29,** 251-256 (1995).

8.  Li X., Cao T., Niu Q., Shi J. R. & Feng J. Coupling the valley degree of freedom to antiferromagnetic order. *Proc. Natl. Acad. Sci. USA* **110,** 3738-3742 (2013).

9.  Sachs B., Wehling T. O., Novoselov K. S., Lichtenstein A. I. & Katsnelson M. I. Ferromagnetic two-dimensional crystals: Single layers of $K_2CuF_4$. *Phys. Rev. B* **88,** 201402 (2013).

10. McGuire M. A., Dixit H., Cooper V. R. & Sales B. C. Coupling of Crystal Structure and Magnetism in the Layered, Ferromagnetic Insulator $CrI_3$. *Chem. Mater.* **27,** 612-620 (2015).

11. Sivadas N., Daniels M. W., Swendsen R. H., Okamoto S. & Xiao D. Magnetic ground state of semiconducting transition-metal trichalcogenide monolayers. *Phys. Rev. B* **91,** 235425 (2015).

12. Du K.-Z., Wang X.-Z., Liu Y., Hu P., Utama M. I. B., Gan C. K., Xiong Q. & Kloc C. Weak Van der Waals Stacking, Wide-Range Band Gap, and Raman Study on Ultrathin Layers of Metal Phosphorus Trichalcogenides. *ACS Nano* **10,** 1738-1743 (2016).

13. May A. F., Calder S., Cantoni C., Cao H. B. & McGuire M. A. Magnetic structure and phase stability of the van der Waals bonded ferromagnet $Fe_{3-x}GeTe_2$. *Phys. Rev. B* **93,** 014411 (2016).

14. Lee S., Choi K. Y., Lee S., Park B. H. & Park J. G. Tunneling transport of mono- and few-layers magnetic van der Waals $MnPS_3$. *Apl Mater.* **4,** 086108 (2016).

15. Lin M. W., Zhuang H. L. L., Yan J. Q., Ward T. Z., Puretzky A. A., Rouleau C. M., Gai Z., Liang L. B., Meunier V., Sumpter B. G., Ganesh P., Kent P. R. C., Geohegan D. B., Mandrus D. G. & Xiao K. Ultrathin nanosheets of $CrSiTe_3$: a semiconducting two-dimensional ferromagnetic material. *J. Mater. Chem. C* **4,** 315-322 (2016).

16. Gong C., Li L., Li Z., Ji H., Stern A., Xia Y., Cao T., Bao W., Wang C., Wang Y., Qiu Z. Q., Cava R. J., Louie S. G., Xia J. & Zhang X. Discovery of intrinsic ferromagnetism in two-dimensional van der Waals crystals. *Nature* **546,** 265-269 (2017).

17. Huang B., Clark G., Navarro-Moratalla E., Klein D. R., Cheng R., Seyler K. L., Zhong D., Schmidgall E., McGuire M. A., Cobden D. H., Yao W., Xiao D., Jarillo-Herrero P. & Xu X. Layer-dependent ferromagnetism in a van der Waals crystal down to the monolayer limit. *Nature* **546,** 270-273 (2017).

18. Zhong D., Seyler K. L., Linpeng X., Cheng R., Sivadas N., Huang B., Schmidgall E., Taniguchi T., Watanabe K., McGuire M. A., Yao W., Xiao D., Fu K.-M. C. & Xu X. Van der Waals engineering of ferromagnetic semiconductor heterostructures for spin and valleytronics. *Sci. Adv.* **3,** e1603113 (2017).

19. Xing W., Chen Y., Odenthal M. P., Zhang X., Yuan W., Su T., Song Q., Wang T., Zhong J., Jia S., Xie X. C., Li Y. & Han W. Electric field effect in multilayer $Cr_2Ge_2Te_6$ : a ferromagnetic 2D material. *2D Mater.* **4,** 024009 (2017).

20. Furdyna J. K. Diluted Magnetic Semiconductors. *J. Appl. Phys.* **64,** R29-R64 (1988).

21. MacDonald A. H., Schiffer P. & Samarth N. Ferromagnetic semiconductors: moving beyond (Ga,Mn)As. *Nat. Mater.* **4,** 195 (2005).

22. Dietl T. & Ohno H. Dilute ferromagnetic semiconductors: Physics and spintronic structures. *Rev. Mod. Phys.* **86,** 187-251 (2014).

23. Wang H., Eyert V. & Schwingenschlögl U. Electronic structure and magnetic ordering of the semiconducting chromium trihalides $CrCl_3$ , $CrBr_3$ , and $CrI_3$. *J. Phys. Condens. Matter* **23,** 116003 (2011).

24. Liu J. Y., Sun Q., Kawazoe Y. & Jena P. Exfoliating biocompatible ferromagnetic Cr-trihalide monolayers. *Phys. Chem. Chem. Phys.* **18,** 8777-8784 (2016).

25. Zhang W. B., Qu Q., Zhua P. & Lam C. H. Robust intrinsic ferromagnetism and half semiconductivity in stable two-dimensional single-layer chromium trihalides. *J. Mater. Chem. C* **3,** 12457-12468 (2015).

26. Li X. X. & Yang J. L. $CrXTe_3$ (X = Si, Ge) nanosheets: two dimensional intrinsic ferromagnetic semiconductors. *J. Mater. Chem. C* **2,** 7071-7076 (2014).

27. Yu W. J., Liu Y., Zhou H., Yin A., Li Z., Huang Y. & Duan X. Highly efficient gate-tunable photocurrent generation in vertical heterostructures of layered materials. *Nat. Nanotech.* **8,** 952 (2013).

28. Britnell L., Ribeiro R. M., Eckmann A., Jalil R., Belle B. D., Mishchenko A., Kim Y.-J., Gorbachev R. V., Georgiou T., Morozov S. V., Grigorenko A. N., Geim A. K., Casiraghi C., Neto A. H. C. & Novoselov K. S. Strong Light-Matter Interactions in Heterostructures of Atomically Thin Films. *Science* **340,** 1311-1314 (2013).

29. Fowler R. H. & Nordheim L. Electron emission in intense electric fields. *Proc. R. Soc. London A* **119,** 173-181 (1928).

30. Lenzlinger M. & Snow E. H. Fowler-Nordheim Tunneling into Thermally Grown $SiO_2$. *J. Appl. Phys.* **40,** 278-283 (1969).

31. Sze S. M. & NG K. K. *Physics of Semiconductor Devices*, Third edn. John Wiley & Sons, Inc.: New Jersey, 2007.

32. Palik E. D. & Furdyna J. K. Infrared and Microwave Magnetoplasma Effects in Semiconductors. *Rep. Prog. Phys.* **33,** 1193-1322 (1970).

33. Song T., Cai X., Tu M. W.-Y., Zhang X., Huang B., Wilson N. P., Seyler K. L., Zhu L., Taniguchi T. & Watanabe K. Giant Tunneling Magnetoresistance in Spin-Filter van der Waals Heterostructures. *Science*, 10.1126/science.aar4851 (2018).

34. Klein D. R., MacNeill D., Lado J. L., Soriano D., Navarro-Moratalla E., Watanabe K., Taniguchi T., Manni S., Canfield P. & Fernández-Rossier J. Probing magnetism in 2D van der Waals crystalline insulators via electron tunneling. *Science*, 10.1126/science.aar3617 (2018).

35. Feng W., Guo G.-Y., Zhou J., Yao Y. & Niu Q. Large magneto-optical Kerr effect in noncollinear antiferromagnets $Mn_3X$ (X = Rh, Ir, Pt). *Phys. Rev. B* **92,** 144426 (2015).

36. Sivadas N., Okamoto S. & Xiao D. Gate-Controllable Magneto-optic Kerr Effect in Layered Collinear Antiferromagnets. *Phys. Rev. Lett.* **117,** 267203 (2016).

37. Worledge D. & Geballe T. Magnetoresistive double spin filter tunnel junction. *J. Appl. Phys.* **88,** 5277-5279 (2000).

38. Miao G.-X., Müller M. & Moodera J. S. Magnetoresistance in Double Spin Filter Tunnel Junctions with Nonmagnetic Electrodes and its Unconventional Bias Dependence. *Phys. Rev. Lett.* **102,** 076601 (2009).

39. Kim H. H., Yang B., Patel T., Sfigakis F., Li C., Tian S., Lei H. & Tsen A. W. One million percent tunnel magnetoresistance in a magnetic van der Waals heterostructure. Preprint at Https://arxiv.org/abs/1804.00028 (2018).

**Acknowledgements**: We gratefully acknowledge A. Ferreira for continuous technical support, D.-K. Ki for fruitful discussions, and A. Ubaldini for early work on the growth of $CrI_3$ crystals. Z.W. thanks C. Handschin for sharing his experience in device fabrication. Z.W., I.G.L. and A.F.M. gratefully acknowledge financial support from the Swiss National Science Foundation, the NCCR QSIT and the EU Graphene Flagship Project. N.U. and M.G. gratefully acknowledge support through an Ambizione fellowship of the Swiss National Science Foundation. M.K. and A.I. acknowledge NCCR QSIT for financial support. K.W. and T.T. acknowledge support from the Elemental Strategy Initiative conducted by the MEXT, Japan and JSPS KAKENHI Grant Numbers JP15K21722.

**Author contributions**: Z.W. fabricated the devices with the collaboration of I.G.L., Z.W. performed the transport measurements and analyzed the data, N.U. performed photoluminescence and photocurrent measurements. M.K. carried out the optical Kerr measurements with help of N.U.; M. K. and A. I. participated in discussion of the Kerr measurements. E.G. grew and characterized the $CrI_3$ crystals. M.G. contributed to the discussion of the possible magnetic states of $CrI_3$ and performed ab-initio calculations to identify when the interlayer coupling is antiferromagnetic. T.T. and K.W. provided hBN crystals. A.F.M. initiated and supervised the project. Z.W., I.G.L., N.U., M.G. and A.F.M. wrote the manuscript with input from all authors. All authors discussed the results.

**Competing financial interests:** The authors declare no competing financial interests.

**Figures:**

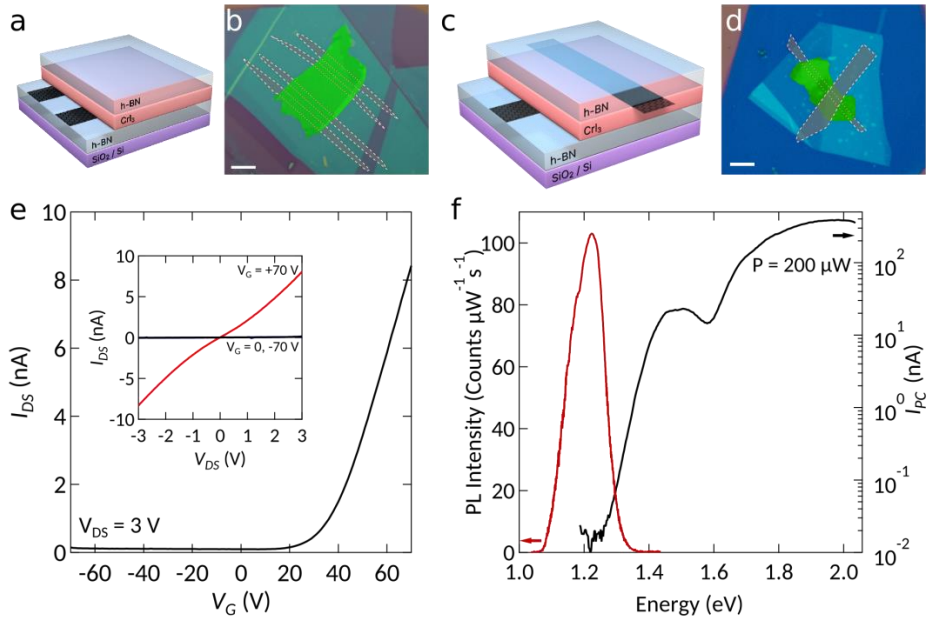

**Figure 1 | Semiconducting characteristics of $CrI_3$.** Scheme (**a**) and false-color optical micrograph (**b**) of a $CrI_3$ field-effect transistor realized using few-layer graphene contacts, encapsulated between hexagonal boron nitride (hBN) crystals. The highly doped Silicon substrate covered by a 285 nm $SiO_2$ layer is used as gate (the scale bar in **b** is 5 μm long). Scheme (**c**) and false-color optical micrograph (**d**) of a heterostructure consisting of bottom and top multilayer graphene contacts attached to an exfoliated $CrI_3$ crystal approximately 7 nm thick (the entire structure is encapsulated between hBN crystals; the scale bar in (**d**) is 5 μm long). **e**, Transfer characteristics of the field-effect transistor shown in (**b**) measured at room temperature with $V_{DS} = 3$ V applied between the two multilayer graphene contacts. The transistor turns on for positive gate voltage indicating electron conduction. The inset shows the source-drain current flowing between the graphene contacts as a function of $V_{DS}$, for three values of gate voltage ($V_G = -70$ V, 0 V, and +70 V). **f**, Dependence of the zero-bias photocurrent (black solid line) and photoluminescence (PL) intensity (red solid line) on the photon excitation energy (data taken at $T = 4$ K, on a device analogous to that shown in **d** with $CrI_3$ of ~10 nm thick).

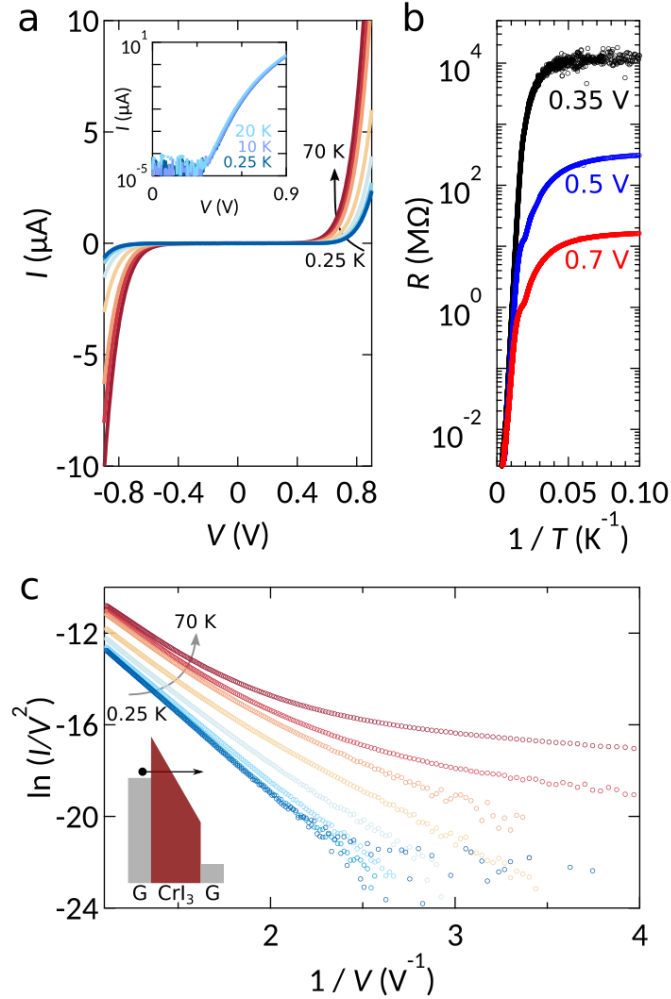

**Figure 2 | Electron tunneling in few-layer CrI$_3$ vertical junctions. a**, Current measured on the device shown in Fig. 1d as a function of bias applied between the graphene contacts for $T$ ranging from 0.25 K to 70 K (the intermediate temperatures are 10 K, 20 K, 30 K, 40 K, 50 K and 60 K). Below $T = 20$ K, the $I$-$V$ curves become temperature-independent as shown in the inset, indicating that transport is determined by tunneling (the overlapping area of the graphene contacts is 4 μm$^2$ and the thickness of the CrI$_3$ layer is approximately 7 nm). **b**, Arrhenius plot of the resistance measured at different bias voltages (0.35 V, 0.5 V and 0.7 V). **c**, In the tunneling regime (i.e., for $T <$ 20 K), $ln(I/V^2)$ is linearly proportional to $1/V$, as expected for Fowler-Nordheim tunneling (charge carriers tunnel into the conduction band through band-gap of CrI$_3$ that is tilted by the applied bias forming a triangular barrier, as illustrated schematically in the inset). The different curves correspond to measurements performed at the same temperatures as in **a**.

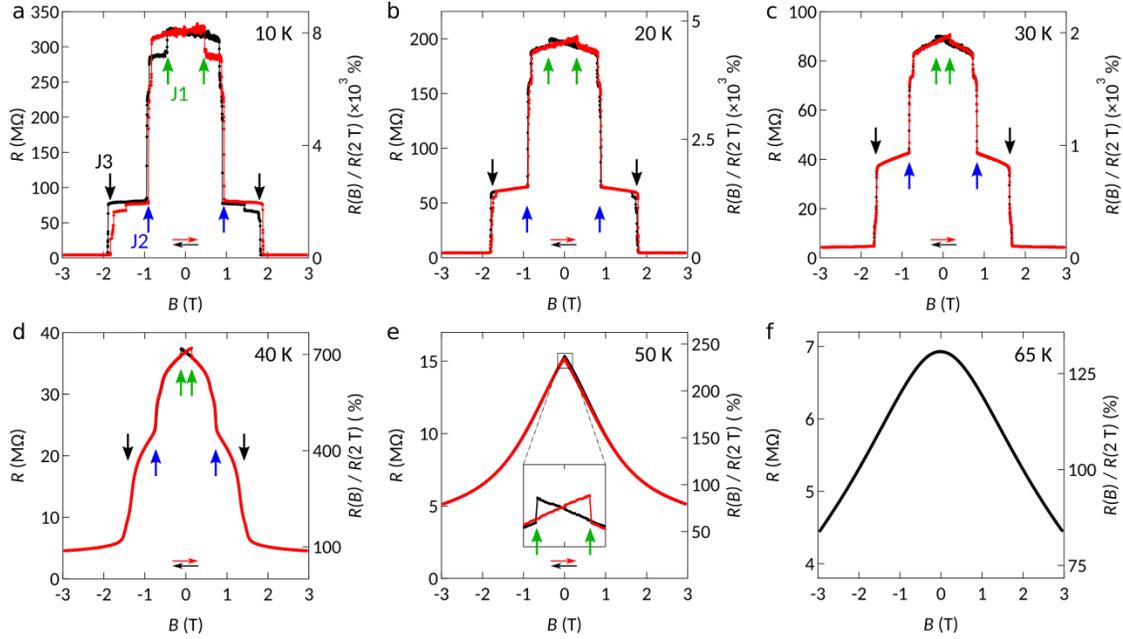

**Figure 3 | Large tunneling magnetoresistance in vertical junctions. a-f** Tunneling resistance (left axis) and resistance ratio $R(B)/R(2\,T)$ (right axis) of the device shown in Fig. 1d, measured at the temperature indicated in each panel (with $V = 0.5$ V and **B** applied perpendicular to the CrI$_3$ layers). The red and black dots correspond to data measured upon sweeping the field in opposite directions as indicated by the horizontal arrows of the corresponding color. The resistance ratio increases upon lowering temperature and reaches 8'000% at 10 K. The arrows of different color point to the magnetoresistance "jumps" that are seen in all devices, irrespective of the thickness of the CrI$_3$ crystal. Jump J1 is always accompanied by hysteresis; at low temperature, jumps J2 and J3 occur in all devices at the same value of the applied magnetic field, irrespective of sweeping direction. All jumps shift to lower field values upon increasing temperature, and disappear above 50 K.

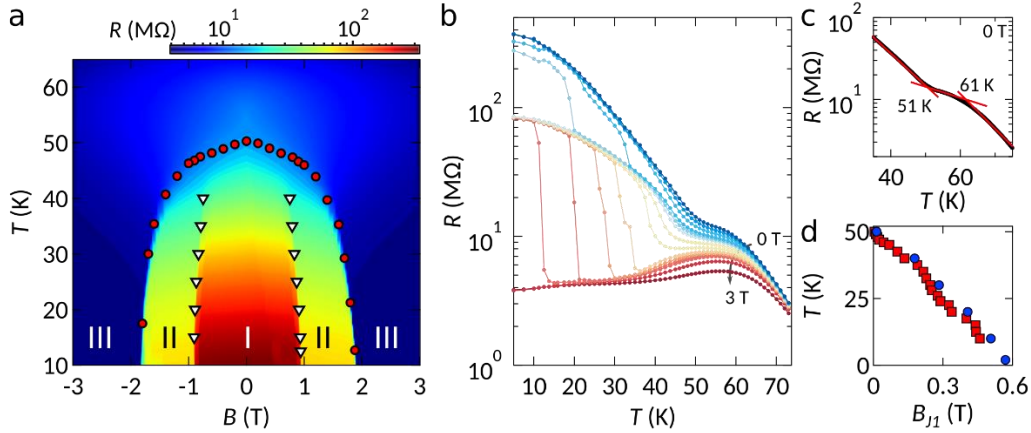

**Figure 4 | Temperature and magnetic field evolution of magnetic states in CrI$_3$. a**, Color plot of the resistance of the device shown in Fig. 1d (in logarithmic scale), as a function of **B** and *T*. At low temperature, three clear plateaus signal the presence of different magnetic states (labeled I, II, III). The white triangles and red circles, obtained from the position of the resistance jumps as described in the text, outline the boundaries of these states, and show that the magnetoresistance features appear at $T \cong 51$ K (i.e., in correspondence of the anomaly seen in the low-field magnetization; see Supplementary Fig. 1c-d). **b**, *T*-dependence of the resistance (in logarithmic scale) at different fixed values of *B*: three different values are attained at low-temperature, corresponding to the different magnetic states of CrI$_3$. **c**, *T*-dependence of the resistance measured at *B*=0 T. The kink at $T \cong 51$ K originates from the evolution of the boundaries between II and III (see **a**); the ferromagnetic transition manifests itself as a kink around 61 K. **d**, Temperature dependence of the position of "jump" J1 (clearly seen in Fig. 3 in linear scale; the logarithmic scale in Fig. 5a makes jump J1 difficult to discern). The blue and red symbols –measured 8 months after each other– demonstrate the excellent reproducibility and stability of encapsulated CrI$_3$ devices.

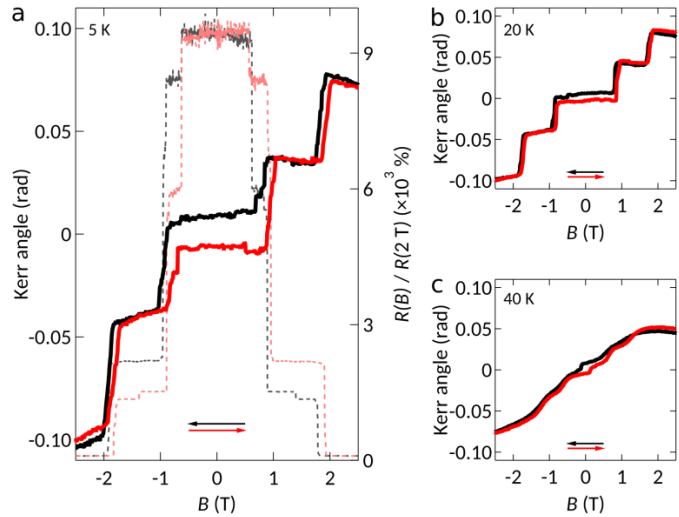

**Figure 5 | Magneto-optical Kerr effect in few-layer CrI$_3$. a**, Comparison between the Kerr angle (solid lines, left axis) and the magnetoresistance (dashed lines; data plotted as resistance ratio $R(B)/R(2\,T)$, right axis) measured on a same device at 5 K. Kerr angle is measured in Faraday geometry with magnetic field applied perpendicular to the plane of CrI$_3$. Black and red curves correspond to sweeping the magnetic field in the direction pointed by the arrows of the corresponding color. The Kerr angle exhibits "jumps" at magnetic field values that coincide perfectly with the jumps observed in the magnetoresistance. **b**, and **c**, Kerr angle measured at 20 K and 40 K, respectively, as a function of magnetic field. The evolution with temperature is virtually identical to that observed for the magnetoresistance (Fig. 3), with features shifting to lower fields and becoming broader as temperature is increased.

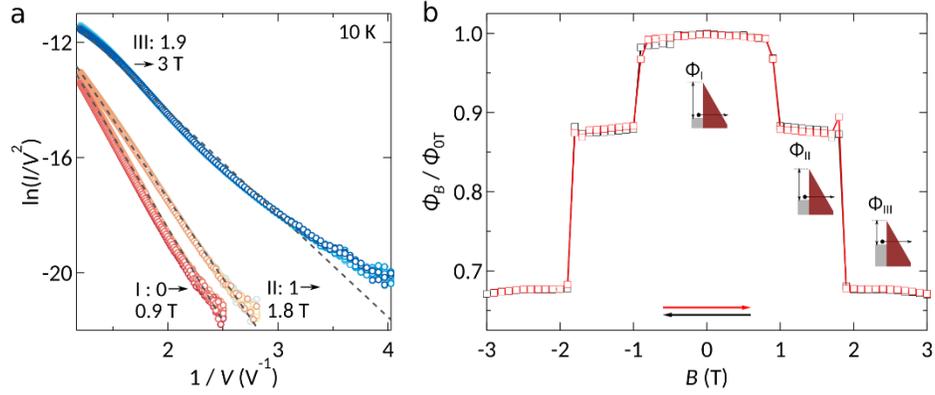

**Figure 6 | Coupling between magnetic state and tunneling resistance of CrI$_3$ a**, Plot of $ln(I/V^2)$ as a function of $1/V$ for **B** ranging between 0 and 3 T showing a nearly linear behavior with a **B**-dependent slope ($T = 10$ K). All data collapse on three different curves. **b**, Magnetic field dependence of the barrier height extracted from the slopes of the curves in **a**, using Eq. (1).

**Very Large Tunneling Magnetoresistance in Layered Magnetic Semiconductor CrI$_3$**

**Supplemental Information**

**Wang** *et al*.

## Supplementary Note 1. Crystal growth and elementary characterization of bulk crystals

Crystals of $CrI_3$ have been grown by the Chemical Vapor Transport method[1]. The elemental precursors Cr (lumps, 99.95% pure, Materials Research SA) and I (crystalline, 99.99+% pure, Alfa Aesar), mixed in the nominal ratio 1:3 with a total mass of 0.3 g, were inserted in a quartz tube inside a glove box filled with 99.9999% Ar. The quartz tube (inner diameter 8 mm) was then tightly connected to a pumping line and evacuated down to ~$10^{-4}$ mbar, with intermediate Ar flushing. The tube was subsequently sealed to a length of ~10 cm and placed horizontally in a tubular furnace with the hot end at 720 °C and the cold end at ~ 640 °C. The thermal treatment lasted 7 days, at the end of which the furnace was switched off and the samples cooled down to room temperature inside the furnace. Shiny, plate-like, dark greyish crystals were found to grow at the cold end of the tube and were easily extracted. Smaller, slightly darker and less shiny ones were also found to grow in the hot zone.

The crystals were characterized by X-ray diffraction in a powder diffractometer (Bragg-Brentano geometry, using a Cu-K X-ray source), which confirmed the $C12/m$ crystal structure, and by electron dispersive X-ray spectroscopy (EDS) in a scanning electron microscope, which confirmed the 1:3 atomic ratio in the final crystals. No traces of $CrI_2$ were found, within the sensitivity of our XRD and EDX probes. Bulk crystals have been proved to remain stable in air for a time long enough for the structural and chemical characterization.

## Supplementary Note 2. Magnetism in bulk CrI3

$CrI_3$ is a layered material (see Supplementary Figure 1a) known to exhibit a transition to an anisotropic ferromagnetic state (Curie temperature $T_c$ =61 K), showing characteristics of an extremely soft ferromagnet, in which the magnetization due to the spins on the Cr atoms is oriented perpendicular to the layers[1-3] (as discussed in Ref 1, the ultra-soft behavior at low magnetic field is likely due to the formation of up and down domains, with domain walls that can move easily through single crystals). Such magnetic behavior is also observed in the magnetization measurements performed on bulk $CrI_3$ crystals grown in our laboratory (Supplementary Figure 1 b-d; see below for details regarding the measurements). In the temperature dependence of the magnetization an anomaly is clearly present at $T \sim 51$ K, i.e., well below the Curie temperature (see insets of Supplementary Figure 1c and d). This anomaly is likely due to a second magnetic transition to a state with a more complex spin configuration, in which the spins are not perfectly aligned in the direction perpendicular to the layers[1].

All magnetic measurements have been performed in a variable temperature MPMS3 SQUID magnetometer (Quantum Design). The horizontal rotator option was used to carefully align the *ab*-plane of $CrI_3$ crystal to be perpendicular or parallel to the applied magnetic field.

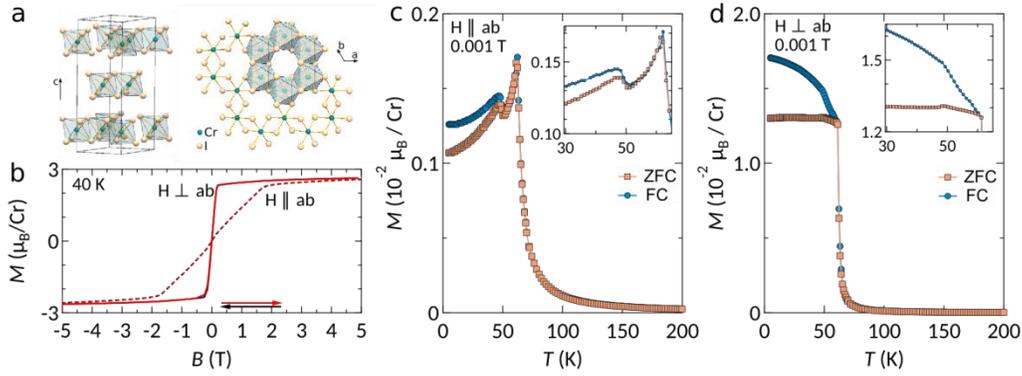

**Supplementary Figure 1 | Magnetism in bulk $CrI_3$. a**, Structure of $CrI_3$ crystal (the scheme represents low temperature equilibrium phase of $CrI_3$). Left: unit cell; right: top view of the *ab* plane. **b**, Anisotropic magnetic field dependence of the magnetization of bulk $CrI_3$ measured at 40 K (solid lines: **B** applied perpendicular to the $CrI_3$ layers; dashed lines: **B** applied parallel to the $CrI_3$ layers), showing the typical behavior of an extremely soft ferromagnet (i.e., the remnant magnetization at **B** = 0 T is vanishingly small). **c-d**, Zero field cooled (orange squares) and field cooled (blue circles) average magnetic moment per Cr atom measured with **B** = 1 mT applied parallel (**c**) or perpendicular (**d**) to *ab*-plane of $CrI_3$, enabling the determination of the Curie temperature, $T_c$ = 61 K. The insets in figures (**c**) and (**d**) zoom-in on the region around 50 K where an anomaly suggestive of an additional phase transition is clearly seen.

## Supplementary Note 3. Sensitivity of $CrI_3$ to atmospheric conditions and encapsulation

Exfoliated thin (<20 nm) $CrI_3$ crystals fully degrade within minutes upon exposure to air. The degradation leads to complete decomposition within 15 minutes, even for relatively thick flakes (i.e., much thicker than monolayers), as shown in Supplementary Figure 2 a-c. To avoid degradation, $CrI_3$ flakes were exfoliated in the inert atmosphere of a glove box (< 0.5 ppm of water and oxygen) and encapsulated in air-stable materials such as graphene or hexagonal boron nitride (hBN), using a dry transfer technique[4]. Supplementary Figure 2 d-f illustrates the situation with successive optical microscope images of different $CrI_3$ flakes on a large hBN crystals: only the $CrI_3$ flakes protected by a top graphene monolayer (black dotted line in Supplementary Figure 2) do not exhibit degradation. The other flakes strongly degrade within minutes.

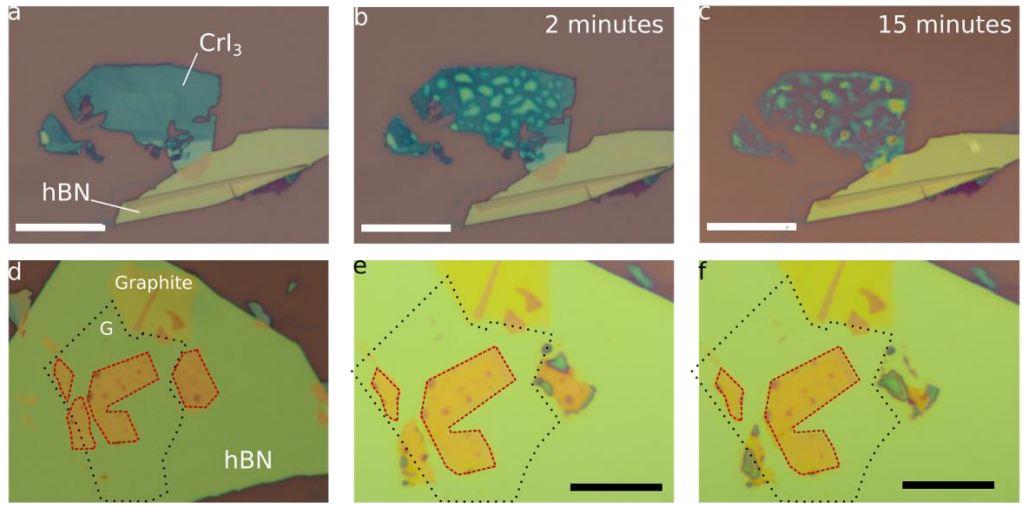

**Supplementary Figure 2 | Degradation of CrI$_3$ thin flakes in air and protection by encapsulation. a-c** Optical microscope images of bare CrI$_3$ thin flake as exfoliated (**a**), after 2 minutes of exposure to ambient (**b**) and after 15 minutes (**c**). The extremely fast degradation process of the material is clearly apparent. **d-f** Optical microscope images of encapsulated CrI$_3$ flakes (whose contour is indicated with the red dashed line) as prepared (**d**), after 2 minutes of exposure to air (**e**) and after 15 minutes (**f**). The images demonstrate that the CrI$_3$ flakes are effectively protected by the bottom hBN crystal and the top graphene monolayer (the contour of the top monolayer graphene used for encapsulation is indicated by the black dotted line). The scale bar in all images is 10 μm long.

In all devices realized to study transport properties (i.e., field effect transistors and vertical junctions), the structures consisting of CrI$_3$ and the multilayer graphene contacts were encapsulated with insulating hBN flakes. Optical microscope images of one device are shown in Supplementary Figure 3 a and b, just after the encapsulation and after metal contact deposition, respectively. No degradation was observed during and after the fabrication process, which included electron beam (e-beam) lithography, PMMA development, reactive ion etching (done to expose the graphene contacts far from the CrI$_3$ crystal, by etching away locally the top hBN layer), e-beam evaporation, and lift-off. Indeed, once properly encapsulated, the devices are stable virtually forever. As an indication, the blue squares and red dots shown in Fig. 6d of the main text were measured 8 months apart without any apparent change in the result. Also, the atomic force microscope image in Supplementary Figure 3c, taken after months of exposure to ambient conditions, shows no sign of degradation.

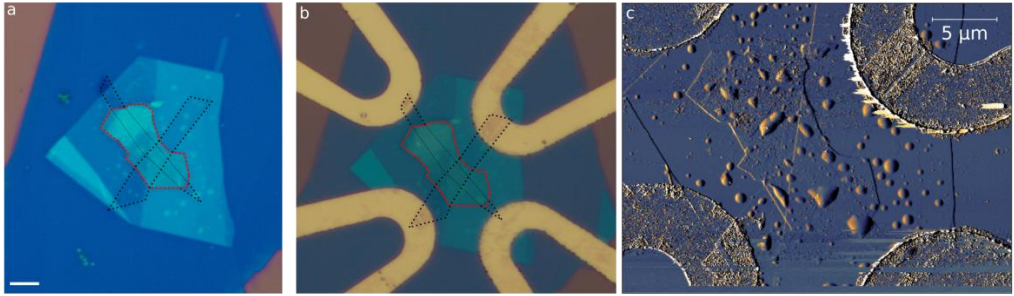

**Supplementary Figure 3 | Stability of an encapsulated CrI$_3$ device. a** Optical microscope image of a vertical Graphene-CrI$_3$-Graphene junction encapsulated between two hBN crystals as extracted from the glove box where the structure was assembeld. The scale bar is 5 μm long. **b** Optical image of the same heterostructure, taken after the multi-layer graphene was side contacted by metal. **c** Atomic force microscopy image of the same device, recorded after the transport measurements. The thickness of the CrI$_3$ flake was determined to be ~ 7 nm ("bubbles" are present in the structure but it is not possible to determine between which of the layers). These images demonstrate that encapsulation of CrI$_3$ is a very robust method against degradation in ambient conditions and withstands standard nano-fabrication techniques.

## Supplementary Note 4. Reproducible magnetoresistance behavior

The key aspects of "vertical" magneto-transport behavior discussed in the main text have been observed in all the samples that we have investigated, as illustrated by the data taken on four different devices shown in Supplementary Figure 4 and 5. In particular, Supplementary Figure 4 and its inset show that jumps J1, J2, J3 (see main text) are present in all devices: jumps J2 and J3 occur at the same values of magnetic field in different devices, whereas jump J1 (see the inset of Supplementary Figure 4) is always accompanied by a hysteresis in the magnetoresistance. It is worth noting that in recent experiments on van der Waals heterostructure of CrI$_3$ and WSe$_2$, a switch in the optical response has been reported to occur at the exact same magnetic field values as the "jump" J2 and J3[5].

The magnitude of the MR depends on the applied bias (as shown in Supplementary Figure5), and typically decreases upon increasing the applied voltage. For each device, the largest magnetoresistance was observed by applying the lowest possible voltage (the smallest voltage is limited by the sensitivity with which we can measure the current), and we found values ranging from 12'500% to 16'600%. In contrast to MR magnitude that was found to depend on bias, the values of the magnetic field at which the MR jumps J2 and J3 occur is the same for all values of applied voltage. Finally, the robust nature of the MR behavior is further illustrated by the temperature evolution of the magnetoresistance with bias voltage of 0.7 V shown in Supplementary Figure6, which exhibits an identical behavior as the data taken at 0.5 V (shown in Supplementary Figure 3).

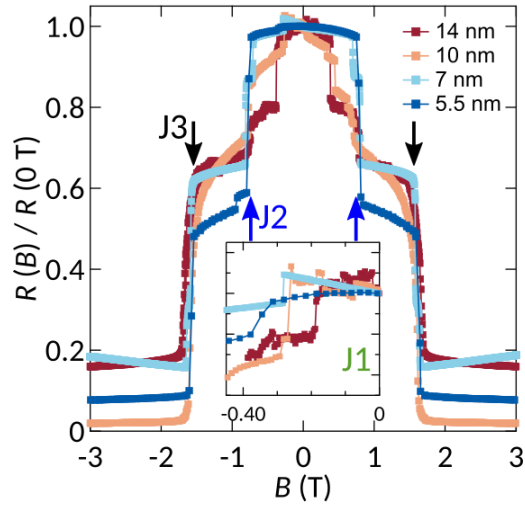

**Supplementary Figure 4 | Magnetoresistance measured in different devices.** Magnetic field dependence of the resistance normalized to zero field for four vertical junctions realized with exfoliated $CrI_3$ crystals of different thickness. The magnetic field is applied perpendicular to $CrI_3$ layers and swept from positive to negative polarity. The applied voltage bias is 3 V for 14 nm device, 1.3 V for 10 nm device, 0.7 V for 7nm device and 0.3 V for 5.5 nm device. The measurements are done at 30 K, a temperature at which very little hysteresis is seen for Jump J2 and J3. All the devices show the resistance jump J2 and J3 at the same field value, +/- 0.8T and +/- 1.5 T, for 30 K. The magnitude of the change in resistance at the jump depends on the applied bias (see Supplementary Figure 5), but the maximum values of magnetoresistance $R(B)/R(2T)$ are comparable in different devices (12 500% for the 5.5 nm device, 16 600% for the 7 nm device and 15 000% for the 10 nm device; for the 14 nm device we have not systematically investigating the magnitude of the MR for different bias). The insert zooms in the data around zero field and demonstrates that the jump J1 is also always present in all devices.

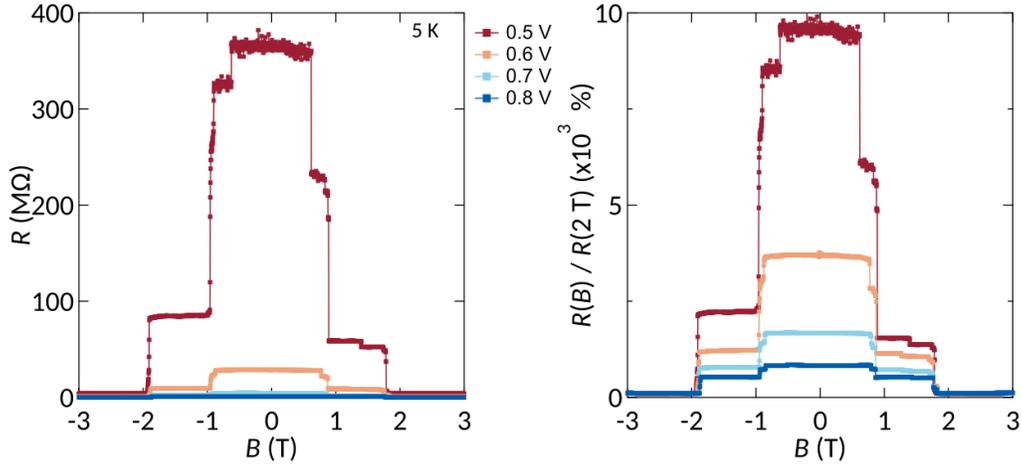

**Supplementary Figure 5 | Magnetoresistance measured at different voltage bias. a-b**, Magnetic field dependence of the resistance (**a**) and resistance ratio *R(B)/ R(2*T*)* (**b**) of the same device whose magneto-resistance is discussed in the main text, measured at 5 K and four different voltages biases (0.5 V, 0.6 V, 0.7 V and 0.8 V, as indicated in the legend). Despite the smaller resistance observed at higher applied bias, due to the reduction of the tunnel barrier width by the higher electric field, all the curves show the presence of the jumps J2 and J3 occurring at the same value of magnetic field.

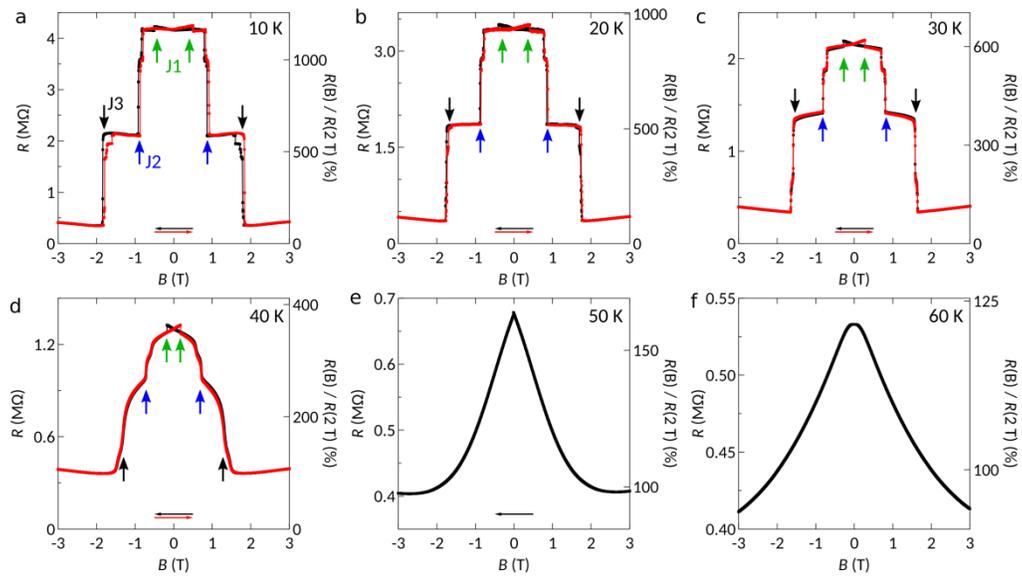

**Supplementary Figure 6 | Temperature evolution of magnetoresistance at a different voltage bias. a**-**f** Magnetoresistance measured with a voltage bias of 0.7 V at 10 K (**a**), 20 K (**b**), 30 K (**c**), 40 K (**d**), 50 K (**e**) and 60 K (**f**). The magnetic field is applied perpendicular to the CrI$_3$ layers. The black and red curves correspond to different sweep directions of the field, as indicated by the horizontal arrows.

**Supplementary Note 5. Magnetoresistance due to an in-plane field**

In the main text we only discussed the magnetoresistance observed upon applying a magnetic field perpendicular to the plane of $CrI_3$. However, a magnetic field applied parallel to the plane also leads to a large magnetoresistance. Supplementary Figure 7 shows the magnetoresistance with **B** applied parallel to the plane measured on a device different from the one discussed in the main text (the thickness of $CrI_3$ crystal in this device is 5.5 nm). The parallel magnetoresistance sets in at a similar temperature and has a comparable magnitude and sensitivity to the applied bias as the one measured when the field is applied perpendicular to the plane. The main differences are that 1) the magnetoresistance exhibits a continuous evolution without discrete steps; 2) the field needed to fully align the spins is somewhat larger than what needed when the field is applied perpendicular to the plane. The value of the resistance measured when the applied field is sufficiently large to align all the spins is approximately the same (although not identical) irrespective of whether the field is applied parallel or perpendicular to the $CrI_3$ layers. These measurements confirm the strongly anisotropic nature of magnetism in $CrI_3$.

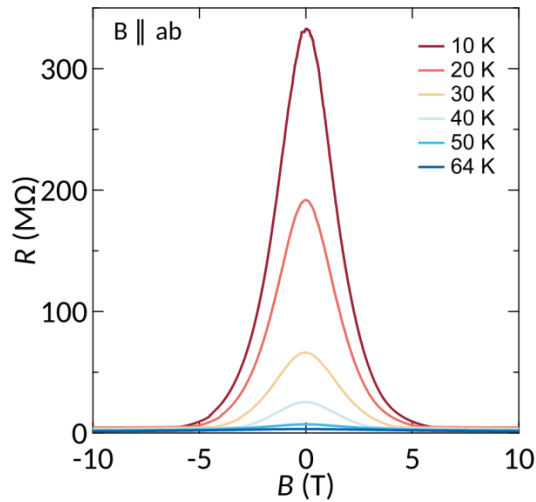

**Supplementary Figure 7| Magnetoresistance with field parallel to the plane.** Temperature evolution of magnetoresistance when the applied magnetic field is parallel to $CrI_3$ layers. In contrast with the case of a perpendicular applied magnetic field, where discrete jumps are observed, the resistance changes smoothly as function of field at all the measured temperatures.

**Supplementary Note 6. Spin configurations with comparable magnetization.**

In the main text we have shown that the magnetoresistance observed in our devices originates from the transition between different magnetic states. In the magnetic field range where the transitions occur, measurements of the bulk magnetization show only very small changes (less that approximately 5%) and no "jumps". This implies that the different magnetic states responsible for the magnetoresistance jumps must have approximately the same magnetization. Supplementary Figure 8a-c are meant to illustrate with simple 1D schemes the qualitative aspects of spin configurations corresponding to different magnetic states having the same –or only minorly different– magnetization. To avoid misunderstandings: the purpose of this discussion is only to illustrate that the constraint that the magnetization does not changes (or changes very little) does not prevent transition between different magnetic states that can lead to different tunneling magnetoresistance values. In fact, since $CrI_3$ is formed by stacking together planes with the spins on the Cr atoms forming a 2D hexagonal lattice in each plane, a large variety of different magnetic states conceptually analogous to those shown in Supplementary Figure 8 can be conceived.

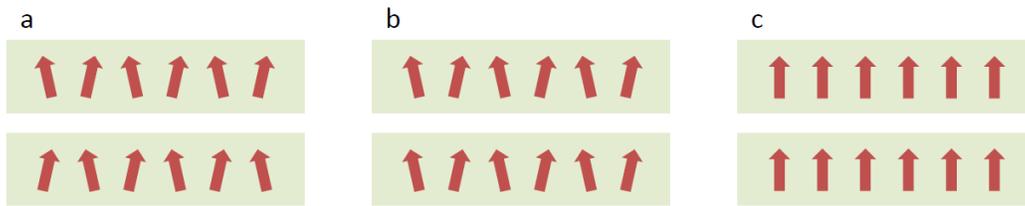

**Supplementary Figure 8| Examples of different magnetic states with almost same magnetization. a-c** spin configurations in two adjacent layers of $CrI_3$ (in a bulk crystal these two layers are repeated periodically in the direction perpendicular to the plane). The spins in **a** and **b** are almost fully polarized perpendicular to the plane, with a small canting due to a spin component exhibiting in-plane antiferromagnetic ordering. The only difference between state (a) and (b) is the spatial alignment of the spin in neighboring planes. The different alignment correspond to different magnetic states, without any different in magnetization. (c) corresponds to the high field spin configuration, with all the spin perfectly aligned perpendicularly to the planes.

## Supplementary Note 7. Exchange coupling from first-principles in bilayer $CrI_3$ with different stacking order

In order to assess possible mechanisms to explain the presumed existence of inter-layer antiferromagnetic coupling in thin $CrI_3$ samples, we investigate here the magnetic ground state of $CrI_3$ bilayers as a function of the stacking order. Indeed, $CrI_3$ undergoes a structural phase transition from a high-temperature monoclinic phase (space group C2/m) to a low-temperature rhombohedral phase (space group R-3). The transition is not sharp and extends over a finite temperature range depending on the cooling history of the sample[1]. The main difference between the two phases consists mainly in the stacking order of the layers, with the structure of the layers themselves remaining almost unaffected across the transition. In the high-temperature phase the layers are displaced along the zigzag direction by approximately a/3, while in the low-temperature phase there is a ABC stacking, with each layer displaced along the armchair direction by $a/\sqrt{3}$. It seems possible that in thin crystals, the phase transition might occur differently, or that the presence of multilayer graphene contacts might affect the transition, so that thin crystals or their outermost layers might display a different stacking order with the respect to the bulk R-3 phase.

To analyze the consequences of such a scenario, we have considered two layers (for simplicity) of $CrI_3$ and have investigated their electronic and magnetic properties for different stacking order using state-of-the-art first-principles density-functional-theory simulations. Starting from an AA stacking, with one layer exactly on top of the other, we have computed the energy of the ferromagnetic (FM) and antiferromagnetic (AFM) configurations for several horizontal displacements between the two layers. In Supplementary Figure 9a we show the FM energy as a function of the displacement (a similar pattern is observed also for AFM). The minimum energy is for AB (or AC) stacking, in agreement with the experimental observation in bulk $CrI_3$. Two additional local minima correspond to the high-temperature (HT) and AA stacking. All other minima are equivalent by symmetry to the previous ones. In Supplementary Figure 9b we instead show the energy of the FM (blue) and AFM (orange) configurations along a path that goes from AB to AA stacking, passing through the HT displacement (see also the white solid line in Supplementary Figure 9a). In agreement with previous calculations for bulk $CrI_3$, with AB stacking the FM configuration is the most stable. One would thus expect free-standing few-layers to display AB(C) stacking and ferromagnetic inter-layer coupling according to DFT simulations. If one allows the layers to have a stacking similar to the high-temperature phase, we instead have an almost perfect degeneracy between FM and AFM configurations, with a mild preference for the latter that might be further enhanced by dipolar interactions. This suggests that if some or all the layers of thin samples are frozen for some reason in the HT stacking, e.g. because of the interaction with the encapsulating layers, we could have AFM interlayer coupling. These considerations, therefore, could explain why thin layers exhibit magnetic properties characteristic of an antiferromagnet (under the assumption that the Kerr effect measurements effectively probe their magnetization) whereas the bulk magnetization exhibits ferromagnetism.

Calculations have been performed using Quantum ESPRESSO[6] with spin-polarized van-der-Waals-compliant functionals[7]. An energy cutoff of 60 Ry for wave functions and 480 Ry for the density have been used, while the Brillouin zone was sampled using a 6x6x1 Monkhorst-Pack grid. Spurious Coulomb interactions between artificial periodic replicas of the bilayers have been removed using a real-space cut-off[8].

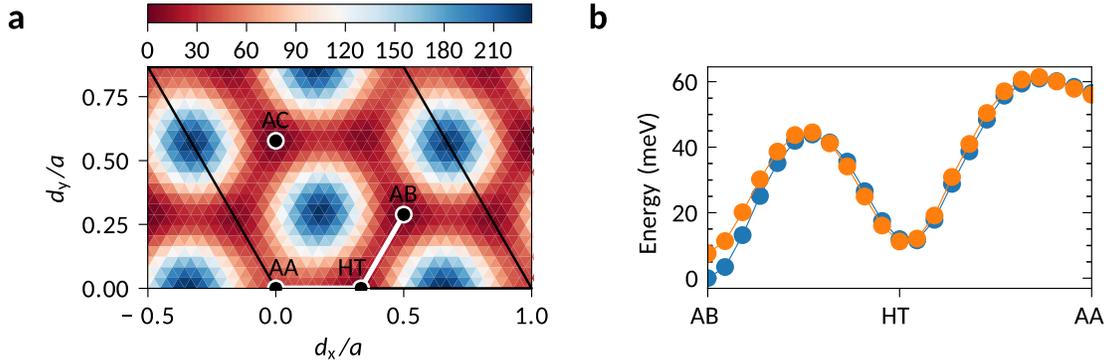

**Supplementary Figure 9| Influence of stacking order on the magnetic state of bilayer CrI$_3$. a.** Colorplot of the energy of bilayer CrI$_3$ in the ferromagnetic configuration as a function of the horizontal displacement $d$ between the layers along the x and y directions in units of the lattice constant $a$. A similar behavior is found also for an antiferromagnetic configuration. The colorbar is the energy difference in meV with respect to the lowest energy AB stacking. The black solid line highlights the primitive cell. **b.** Energy of the ferromagnetic (blue) and antiferromagnetic (orange) configurations as a function of the displacement between the two layers in CrI$_3$ along a path going from the lowest-energy AB stacking to AA staking, passing through the high-temperature (HT) order (see also the white solid line in panel **a**).

## Supplementary Note 8. Tunneling current in the presence of a non-uniform and spin-dependent potential barrier

In the main text, we have shown that the measured current through CrI$_3$ is due to tunneling at low temperatures and follows the Fowler-Nordheim (FN) law reported in Eq. (1). The FN behavior survives also in the presence of a vertical magnetic field, although the current changes in a step-like fashion as a function of the applied B-field, giving rise to the large magnetoresistance discussed in the main text. We have argued that this effect can be accounted for by a variation in the effective barrier height entering the FN expression for the tunneling current (see Fig. 6) that is different in the different magnetic states of CrI$_3$. Here we show that a change in magnetoresistance due to a switch from an antiferromagnetic to a ferromagnetic interlayer ordering in the magnetization does also exhibit a similar behavior, compatible with the FN expression of tunneling and different barrier heights in the ferro- and antiferromagnetic configurations.

To this end we generalize the original derivation by Fowler and Nordheim[9] by considering a potential barrier that even at zero bias is not spatially uniform and depends on the spin orientation. For simplicity, we consider the barrier to be constant within each layer and equal to $\phi_B^P$ for the spin component parallel to the magnetization of the layer and $\phi_B^{AP}$ when for the spin component is antiparallel to the local magnetization. This means that in the ferromagnetic configuration at zero bias electrons with a given spin see a constant barrier height $\phi_B^P$, while electrons with opposite spin feel a barrier with height $\phi_B^{AP} > \phi_B^P$. On the contrary, in the antiferromagnetic configuration, both spins experience a barrier alternating between $\phi_B^P$ and $\phi_B^{AP}$ in neighboring layers, although with opposite phase for the two spin components. At finite bias the barrier is tilted by the electric field as shown in the insets of Supplementary Figure 10, where solid and dashed lines are used to distinguish different spin components and the blue (orange) color refers to a ferromagnetic (antiferromagnetic) configuration for four layers. The tunneling current in the zero-temperature limit is computed as[10]

$$I(V) \propto \int_{E_F-eV}^{E_F} (E_F - E) \sum_\sigma T_\sigma(E)\, dE$$

where $E_F$ is the Fermi energy (assumed to be the same in both electrodes), $V$ is the applied bias, and $T_\sigma(E)$ is the tunneling probability for spin-$\sigma$ electrons. Within the WKB approximation $T_\sigma(E)$ is given by

$$T_\sigma(E) \propto \exp\left\{-2\int_0^d \sqrt{\frac{2m^*}{\hbar^2}[V_\sigma(z) - E]}\, dz\right\}$$

where $d$ is the overall thickness of the $CrI_3$ barrier. The integral is limited to regions where $V_\sigma(z) > E$, while further scattering or interference events are neglected.

In Supplementary Figure 10 we report on our results for the current in the ferromagnetic (blue) and antiferromagnetic configurations for four layers of $CrI_3$. We plot $I/V^2$ as a function of the inverse bias $1/V$ in a semi-logarithmic scale to emphasize that in both cases the current follows the FN behavior. In the ferromagnetic case, the current is not only larger in magnitude but also has a smaller FN slope with respect to the antiferromagnetic case. As mentioned in the main text, the effect can be rationalized in terms of a smaller effective barrier in the ferromagnetic case. Indeed, in this case only one spin component (parallel to the magnetization) contributes to the current with barrier height equal to $\phi_B^P$. On the contrary, in the antiferromagnetic case, both spin components contribute to the current but feel a larger effective barrier in between $\phi_B^P$ and $\phi_B^{AP}$. This is in perfect agreement with experiments if we assume that the zero-field spin configuration is antiferromagnetic, while it is ferromagnetic at larger **B**-fields. These results provide clear evidence that interpreting the magnetoresistance in terms of a change in effective barrier height is a physically reasonable point of view, which can apply also if the magnetic states are more complex than the simple ferromagnetic and antiferromagnetic interlayer configurations.

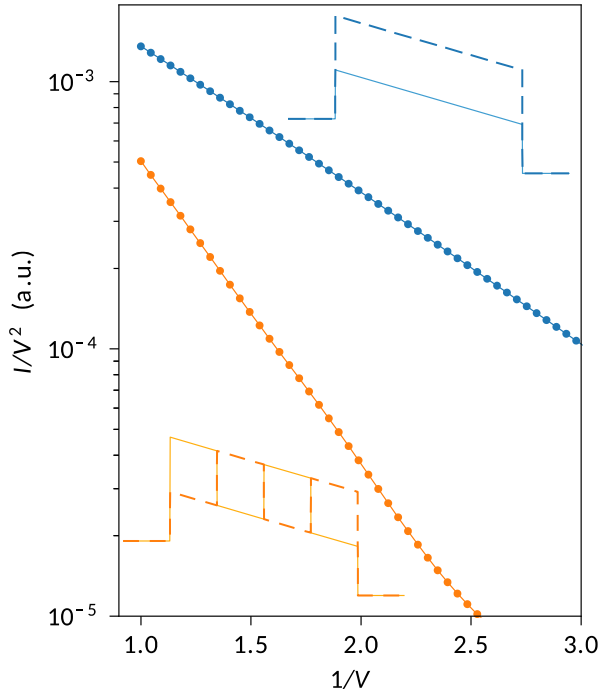

**Supplementary Figure 10| Fowler-Nordheim tunneling current in the presence of a non-uniform and spin-dependent barrier.** Tunneling current $I$ computed as a function of the bias voltage $V$ and plotted as $I/V^2$ versus $1/V$ in a semi-logarithmic scale to emphasize the FN behavior. The calculation has been performed considering four layers of CrI$_3$ with an overall thickness $d = 2.64$ nm. The barrier heights have been set to $\phi_B^P = 0.18$ eV and $\phi_B^{AP} = 0.38$ eV relative to the Fermi energy $E_F = 1.5$ eV in both electrodes at zero bias. The insets show the profile of the barrier in ferromagnetic (top, blue) and antiferromagnetic configuration (bottom, orange), where solid and dashed lines distinguish opposite spin components. In the ferromagnetic case the tunneling current is larger in magnitude and has a smaller FN slope than in the antiferromagnetic case, signaling a smaller effective barrier height.

## Supplementary References:

1.  McGuire M. A., Dixit H., Cooper V. R. & Sales B. C. Coupling of Crystal Structure and Magnetism in the Layered, Ferromagnetic Insulator CrI$_3$. *Chem. Mater.* **27,** 612-620 (2015).

2.  Dillon J. F. & Olson C. E. Magnetization Resonance and Optical Properties of Ferromagnet CrI$_3$. *J. Appl. Phys.* **36,** 1259-1260 (1965).

3.  Dillon J. F., Kamimura H. & Remeika J. P. Magneto-optical properties of ferromagnetic chromium trihalides. *J. Phys. Chem. Solids* **27,** 1531-1549 (1966).

4. Wang L., Meric I., Huang P. Y., Gao Q., Gao Y., Tran H., Taniguchi T., Watanabe K., Campos L. M., Muller D. A., Guo J., Kim P., Hone J., Shepard K. L. & Dean C. R. One-Dimensional Electrical Contact to a Two-Dimensional Material. *Science* **342,** 614-617 (2013).

5. Zhong D., Seyler K. L., Linpeng X., Cheng R., Sivadas N., Huang B., Schmidgall E., Taniguchi T., Watanabe K., McGuire M. A., Yao W., Xiao D., Fu K.-M. C. & Xu X. Van der Waals engineering of ferromagnetic semiconductor heterostructures for spin and valleytronics. *Sci. Adv.* **3,** e1603113 (2017).

6. Giannozzi P., Baroni S., Bonini N., Calandra M., Car R., Cavazzoni C., Ceresoli D., Chiarotti G. L., Cococcioni M., Dabo I., Dal Corso A., de Gironcoli S., Fabris S., Fratesi G., Gebauer R., Gerstmann U., Gougoussis C., Kokalj A., Lazzeri M., Martin-Samos L., Marzari N., Mauri F., Mazzarello R., Paolini S., Pasquarello A., Paulatto L., Sbraccia C., Scandolo S., Sclauzero G., Seitsonen A. P., Smogunov A., Umari P. & Wentzcovitch R. M. QUANTUM ESPRESSO: a modular and open-source software project for quantum simulations of materials. *Journal of Physics-Condensed Matter* **21,** (2009).

7. Thonhauser T., Zuluaga S., Arter C. A., Berland K., Schröder E. & Hyldgaard P. Spin Signature of Nonlocal Correlation Binding in Metal-Organic Frameworks. *Phys. Rev. Lett.* **115,** 136402 (2015).

8. Sohier T., Calandra M. & Mauri F. Density functional perturbation theory for gated two-dimensional heterostructures: Theoretical developments and application to flexural phonons in graphene. *Phys. Rev. B* **96,** 075448 (2017).

9. Fowler R. H. & Nordheim L. Electron emission in intense electric fields. *Proc. R. Soc. London A* **119,** 173-181 (1928).

10. Simmons J. G. Generalized formula for the electric tunnel effect between similar electrodes separated by a thin insulating film. *J. Appl. Phys.* **34,** 1793-1803 (1963).